\documentclass[11pt]{article}

\usepackage[margin=1.1in]{geometry}
\usepackage{graphicx}
\usepackage{tikz}
\usetikzlibrary{arrows.meta}

\usepackage{amssymb}
\newcommand{\zz}{{\mathbb Z}}
\newcommand{\cc}{{\mathbb C}}
\newcommand{\bfa}{{\bf a}}
\newcommand{\bfc}{{\bf c}}
\newcommand{\bfe}{{\bf e}}
\newcommand{\bff}{{\bf f}}
\newcommand{\bfg}{{\bf g}}
\newcommand{\bfh}{{\bf h}}
\newcommand{\x}{{\bf x}}
\newcommand{\zero}{{\bf 0}}

\newtheorem{theorem}{Theorem}[section]
\newtheorem{proposition}[theorem]{Proposition}
\newtheorem{corollary}[theorem]{Corollary}
\newtheorem{lemma}[theorem]{Lemma}

\begin{document}

\title{Locating the Closest Singularity in a Polynomial 
Homotopy\thanks{Supported by the National Science Foundation 
under grant DMS 1854513.}}

\author{Jan Verschelde\thanks{University of Illinois at Chicago,
Department of Mathematics, Statistics, and Computer Science,
851 S. Morgan St. (m/c 249), Chicago, IL 60607-7045.
Email: {\tt janv@uic.edu}, URL: {\tt http://www.math.uic.edu/$\sim$jan}.}
\and Kylash Viswanathan\thanks{University of Illinois at Chicago,
Department of Mathematics, Statistics, and Computer Science,
851 S. Morgan St. (m/c 249), Chicago, IL 60607-7045.
Email: {\tt kviswa5@uic.edu}.}}

\date{25 June 2022}

\maketitle

\begin{abstract}
A polynomial homotopy is a family of polynomial systems, 
where the systems in the family depend on one parameter.
If for one value of the parameter we know a regular solution,
then what is the nearest value of the parameter for which the 
solution in the polynomial homotopy is singular?
For this problem we apply the ratio theorem of Fabry.
Richardson extrapolation is effective to accelerate the convergence
of the ratios of the coefficients of the series expansions of the
solution paths defined by the homotopy.
For numerical stability, we recondition the homotopy.
To compute the coefficients of the series
we propose the quaternion Fourier transform.
We locate the closest singularity computing at a regular solution,
avoiding numerical difficulties near a singularity.
\end{abstract}

\section{Introduction}

Polynomial homotopies define the deformation of polynomial systems,
from systems with known solutions into systems that must be solved.
We call a solution {\em regular} if the matrix of all partial
derivatives evaluated at the solution has full rank,
otherwise the solution is {\em singular}.
We aim to locate
the nearest singularity starting at a regular solution.
Applying the ratio theorem of Fabry, we can detect singular points 
based on the coefficients of the Taylor series.

\begin{theorem} [the ratio theorem of Fabry~\cite{Fab1896}] \label{thmFabry}
If for the series
$x(t) = c_0 + c_1 t + c_2 t^2 + \cdots + c_n t^n + c_{n+1} t^{n+1} + \cdots$,
we have
$\displaystyle \lim_{n \rightarrow \infty} c_n/c_{n+1} = z$, then
\begin{itemize}
\item $z$ is a singular point of the series, and 
\item it lies on the boundary of the circle of convergence of the series.
\end{itemize}
Then the radius of this circle is less than $|z|$.
\end{theorem}

While the proof of the theorem
would take us deep into complex analysis~\cite[Chapter~XI]{Die57},
one can immediately verify that
the ratio $c_n/c_{n+1}$ is the pole of
Pad\'{e} approximants (\cite{BG96}, \cite{Sue02}) of degrees~$[n/1]$,
where $n$ is the degree of the numerator, with linear denominator.

The ratio theorem of Fabry provides a radar to detect singularities
in an adaptive step size control for continuation methods,
as introduced in~\cite{TVV20a} (with a parallel implementation
in~\cite{TVV20b}) and reproduced by~\cite{Tim21}.
Earlier applications of Pad\'{e} approximants in deformation
methods appeared in~\cite{JMSW09}, in a symbolic context,
and in~\cite{SC87} in a numerical setting.
Empirically, in the plain application of this ratio theorem,
already relatively few terms in the series appear to be sufficient
to take nearby singularities into account.

The problem considered in this paper can be stated as follows.
How many terms in the Taylor series do we need to locate
the closest singularity with eight decimal places of accuracy?
Answering this question exactly is not possible because of constants
which differ for each series, but we can provide information about
the order of the number of terms, e.g.: tens or hundreds.

We show that Richardson extrapolation
(see~\cite{Bre80} for a general formalism)
effectively solves our problem.
On monomial homotopies (defined in the next section),
we can separate our problem from the required accuracy of the
coefficients of the Taylor series.
On examples, at 64 terms of the series, we obtain eight decimal places of
accuracy in the location of the radius of convergence.
In the third section, the justification for this successful
application of Richardson extrapolation is proven.  
This is the first contribution of this paper.

The second contribution of this paper is the introduction
of the quaternion Fourier transform~\cite{ELS14}, \cite{SLS08} 
to compute the coefficients of the series.
If we want to locate a singularity to full double precision,
then, on examples, it appears that 512 terms in the series are needed.
The Fast Fourier Transform scales well.

In the fifth section, we consider the application of Richardson
extrapolation in an end game, when the path tracker approaches
an isolated singular solution at the end of the path.
Power series methods for singular solutions in~\cite{MSW92}
introduced the concept of the {\em end game operation range}.
In this range, the continuation parameter has values for which 
the Puiseux series expansions are valid and where the numerical 
condition numbers still allow to compute
sufficiently accurate approximations of the points on the path.
In fixed precision, this range may be empty.
Using multiple double precision for ill-conditioned problems is 
wasteful due to the slow convergence of Newton's method.
For homotopies with a random complex gamma constant,
we introduce the notion of the last pole.
With this last pole, we {\em recondition} the homotopy 
with a shift and stretch transformation.


The new methods are illustrated in section six.
Deflation restores the quadratic convergence of Newton's method
for an isolated singular solution, of multiplicity~$\mu$.
While~\cite{LVZ06} proves that $\mu$ is the upper bound on the
number of deflation steps, the numerical decision to apply deflation
is left to a singular value decomposition of the Jacobian matrix,
which may not always be reliable enough.
Although deflation has been addressed by many
(e.g.~\cite{BL21}, \cite{CDW20}, \cite{DZ05}, \cite{DLZ11}, \cite{GLSY07}, 
\cite{HJLZ20}, \cite{HMS17}, \cite{LZ12,LZ14}, \cite{Oji87a}),
the question on when to deflate is an open problem.

\newpage

\section{Monomial Homotopies}

The examples of the homotopies in this section 
have only one singularity.

A {\em monomial homotopy} is defined by an exponent matrix
$A \in \zz^{n \times n}$ and an $n$-dimensional coefficient vector $\bfc(t)$
of invertible power series:
\begin{equation}
   \bfh(\x, t) = \x^A - \bfc(t) = \zero, \quad
\end{equation}
with $\x = (x_1, x_2, \ldots, x_n)$, and
the multi-index notation
\begin{equation}
   \bfa_j = (a_{1,j}, a_{2, j}, \ldots, a_{n, j}), \quad 
   \x^{\bfa_j} = x_1^{a_{1,j}}  x_2^{a_{2, j}} \cdots x_n^{a_{n,j}},
\end{equation}
where $\bfa_j$ is the $j$th column of the matrix~$A$.

For any specific value for $t$, the system $\bfh(\x, t) = \zero$
reduces to a system with exactly two monomials in every equation.
The solving of such a system happens via a unimodular coordinate
transformation defined by the Hermite normal form of~$A$.
Singular solutions can occur only when $\bfc(t) = \zero$,
only for specific values of~$t$.
While monomial homotopies have thus no direct practical use,
they provide good test cases to experiment with algorithms
and to introduce new ideas.

\subsection{A Square Root Homotopy}

\begin{figure}[hbt]
\centerline{\includegraphics[height=6cm]{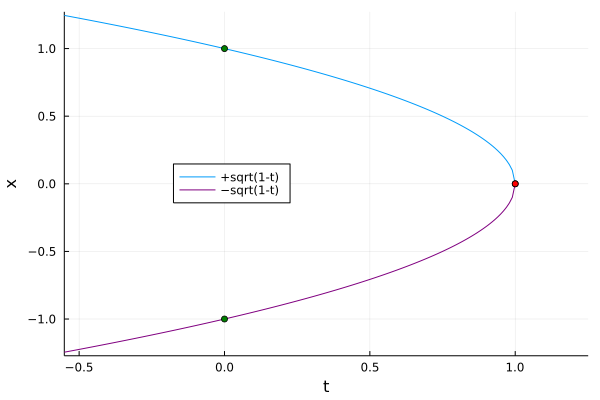}}
\caption{Starting at $x = \pm 1$, the two paths converge
to $x = 0$, as $t$ moves from~0 to~1.}
\label{figsqrthomotopy}
\end{figure}

The simplest example of a monomial homotopy is
\begin{equation}
   x^2 - 1 + t = 0, \quad \mbox{with solution} \quad x(t) = \pm \sqrt{1 - t}.
\end{equation}
The two paths defined by this homotopy are shown 
in Figure~\ref{figsqrthomotopy}.

At~$t = 1$, the two paths coincide at a double point.
Our problem is to predict for which value of~$t$ this singularity happens
{\em without computing $x(t)$ for $t \approx 1$.}

In the development of the solution $\displaystyle x(t) = \sqrt{1 - t}$
in a Taylor series about $t = 0$, let $c_n$ be the coefficient of~$t^n$.
Then the application of the ratio theorem of Fabry gives
\begin{equation} \label{eqSqrtFabry}
    \frac{c_n}{c_{n+1}}
     = \frac{2(n+1)}{ 2n - 1} =: f(n), \quad
    \lim_{n \rightarrow \infty} f(n) = 1.
\end{equation}
As the limit of the ratios equals one, we can predict the location
of the singularity, already at the series development at~$t = 0$.
The main problem is the slow convergence of the series.
Table~\ref{tabsqrthomerrors} illustrates that in order to gain
one extra bit of accuracy, we must double the value of~$n$.

\begin{table}[hbt]
\begin{center}
\caption{$f(n) = \frac{2(n+1)}{ 2n - 1}$ converges slowly to one.
The error column lists $|f(n) - 1|$.
The last column is the ratio of two consecutive errors.
As $n$ doubles, the error is cut in half.}
\begin{tabular}{c||c|c|c}
   $n$ & $f(n)$ & error & error ratio \\ \hline \hline
   2 &  2.00000000000000 & $1.00\mbox{E}\!+\!00$ & \\
   4 &  1.42857142857143 & $4.29\mbox{E}\!-\!01$ & $2.3333\mbox{E}\!+\!00$ \\
   8 &  1.20000000000000 & $2.00\mbox{E}\!-\!01$ & $2.1429\mbox{E}\!+\!00$ \\
  16 &  1.09677419354839 & $9.68\mbox{E}\!-\!02$ & $2.0667\mbox{E}\!+\!00$ \\
  32 &  1.04761904761905 & $4.76\mbox{E}\!-\!02$ & $2.0323\mbox{E}\!+\!00$ \\
  64 &  1.02362204724409 & $2.36\mbox{E}\!-\!02$ & $2.0159\mbox{E}\!+\!00$ \\
 128 &  1.01176470588235 & $1.18\mbox{E}\!-\!02$ & $2.0079\mbox{E}\!+\!00$ \\
 256 &  1.00587084148728 & $5.87\mbox{E}\!-\!03$ & $2.0039\mbox{E}\!+\!00$ \\
 512 &  1.00293255131965 & $2.93\mbox{E}\!-\!03$ & $2.0020\mbox{E}\!+\!00$
\end{tabular}
\label{tabsqrthomerrors}
\end{center}
\end{table}

Observe we can rewrite $f(n$) of~(\ref{eqSqrtFabry}) as
\begin{eqnarray}
     f(n) & = & \frac{2(n+1)}{2n - 1} = \frac{2n - 1 + 3}{2n -1}
          = 1 + \frac{3}{2n - 1} 
          = 1 + \frac{3}{2n} \left( \frac{1}{1 - \frac{1}{2n}} \right) \\
          & = & 1 + \frac{3}{2n} 
\left( 1 + \frac{1}{2n} 
+ \left( \frac{1}{2n} \right)^2
+ \left( \frac{1}{2n} \right)^3 + \cdots \right).
\end{eqnarray}

As shown in section~3,
$f(n)$ has an asymptotic expansion of the form
\begin{equation}
   f(n) = 1 
+ \gamma_1 \left( \frac{1}{n} \right)
+ \gamma_2 \left( \frac{1}{n} \right)^2
+ \gamma_3 \left( \frac{1}{n} \right)^3 + \cdots
\end{equation}
for some coefficients $\gamma_1$, $\gamma_2$, $\gamma_3$, $\ldots$.
If we double the value for $n$, we have
\begin{equation}
   f(2n) = 1 
+ \gamma_1 \left( \frac{1}{2n} \right)
+ \gamma_2 \left( \frac{1}{2n} \right)^2
+ \gamma_3 \left( \frac{1}{2n} \right)^3 + \cdots
\end{equation}
and then we eliminate $\gamma_1$ via a linear combination:
\begin{equation}
   2 f(2n) - f(n) = 1
+ 2 \gamma_2 \left( \frac{1}{2n} \right)^2
- \gamma_2 \left( \frac{1}{n} \right)^2
+ 2 \gamma_3 \left( \frac{1}{2n} \right)^3
- \gamma_3 \left( \frac{1}{n} \right)^3 + \cdots
\end{equation}
which results in an approximation with error $O(1/n^2)$.

This regular ratio of two consecutive errors allows for an
effective application of Richardson extrapolation.
The input to Richardson extrapolation are the values
$f(2)$, $f(4)$, $f(8)$, $\ldots$, $f(2^N)$.
The output is $R_{i,j}$, the triangular table of extrapolated values.
Then the extrapolation proceeds as follows:

\begin{enumerate}
\item The first column: $R_{i, 1} = f(2^i)$, for $i=1,2,3, \ldots, N$.

\item The next columns in the table are computed via
\begin{equation} \label{eqRichardson}
   R_{i,j} = \frac{2^{i-j+1} R_{i,j-1} - R_{j-1,j-1}}{2^{i-j+1} - 1},
\end{equation}
for $i=i,i+1, \ldots, N$ and for $j=2, 3, \ldots, N$.
\end{enumerate}
Table~\ref{tabsqrthomRichardson} shows the errors $|R_{i,j} - 1|$
of the extrapolated values.
Looking at the diagonal of Table~\ref{tabsqrthomRichardson},
we see that we gain about two decimal places of accuracy
at each doubling of~$n$.

\begin{table}[hbt]
\begin{center}
\caption{Errors of Richardson extrapolation.
The column $E_0$ is the error column of Table~\ref{tabsqrthomerrors}.
The column $E_j$ is the error obtained from extrapolating $j$ times,
applying formula~(\ref{eqRichardson}).
At $n = 64$ we have 8 correct decimal places and at $n=512$,
the full machine precision is attained.}
{\small
\begin{tabular}{c||c|c|c|c|c|c|c|c|c}
  $n$ & $E_0$ &  $E_1$  &  $E_2$  &  $E_3$  &  $E_4$  &  $E_5$  &  $E_6$
      &  $E_7$  &  $E_8$ \\ \hline \hline
  2 & $1.0\mbox{E}\!+\!0$ &         &         &         &     
    &         &         &         &        \\
  4 & $4.3\mbox{E}\!-\!1$ & $1.4\mbox{E}\!-\!1$
    &   &   &   &   &   &   &    \\
  8 & $2.0\mbox{E}\!-\!1$ & $6.7\mbox{E}\!-\!2$
    & $9.5\mbox{E}\!-\!3$ &  &  &  &  &  &     \\
 16 & $9.7\mbox{E}\!-\!2$ & $3.2\mbox{E}\!-\!2$
    & $4.6\mbox{E}\!-\!3$ & $3.1\mbox{E}\!-\!4$
    &         &         &         &         &        \\
 32 & $4.8\mbox{E}\!-\!2$ & $1.6\mbox{E}\!-\!2$
    & $2.3\mbox{E}\!-\!3$ & $1.5\mbox{E}\!-\!4$
    & $4.9\mbox{E}\!-\!6$ &  &  &  &        \\
 64 & $2.4\mbox{E}\!-\!2$ & $7.9\mbox{E}\!-\!3$
    & $1.1\mbox{E}\!-\!3$ & $7.5\mbox{E}\!-\!5$
    & $2.4\mbox{E}\!-\!6$ & $3.8\mbox{E}\!-\!8$
    &         &         &        \\
128 & $1.2\mbox{E}\!-\!2$ & $3.9\mbox{E}\!-\!3$
    & $5.6\mbox{E}\!-\!4$ & $3.7\mbox{E}\!-\!5$
    & $1.2\mbox{E}\!-\!6$ & $1.9\mbox{E}\!-\!8$
    & $1.5\mbox{E}\!-\!10$ &         &        \\
256 & $5.9\mbox{E}\!-\!3$ & $2.0\mbox{E}\!-\!3$
    & $2.8\mbox{E}\!-\!4$ & $1.9\mbox{E}\!-\!5$
    & $6.0\mbox{E}\!-\!7$ & $9.5\mbox{E}\!-\!9$
    & $7.5\mbox{E}\!-\!11$ & $2.9\mbox{E}\!-\!13$ &  \\
512 & $2.9\mbox{E}\!-\!3$ & $9.8\mbox{E}\!-\!4$
    & $1.4\mbox{E}\!-\!4$ & $9.3\mbox{E}\!-\!6$
    & $3.0\mbox{E}\!-\!7$ & $4.8\mbox{E}\!-\!9$
    & $3.8\mbox{E}\!-\!11$ & $1.5\mbox{E}\!-\!13$
    & $4.4\mbox{E}\!-\!16$
\end{tabular}
}
\end{center}
\label{tabsqrthomRichardson}
\end{table}

\subsection{Two Paths Ending in a Cusp}

\begin{figure}[hbt]
\centerline{\includegraphics[height=6cm]{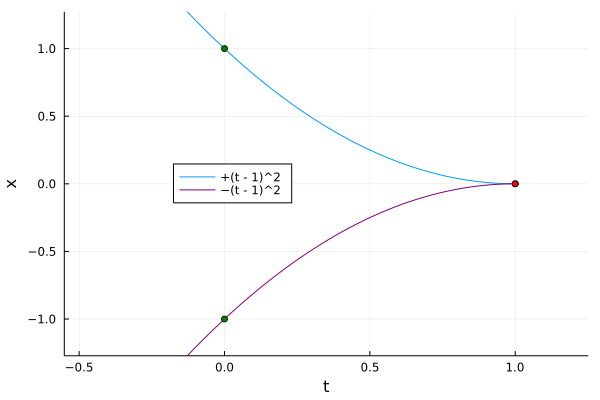}}
\caption{Starting at $x = \pm 1$, the two paths converge
to $x = 0$, as $t$ moves from~0 to~1.}
\label{figsquarehomotopy}
\end{figure}

Figure~\ref{figsquarehomotopy} is an example of a situation
not covered by Theorem~\ref{thmFabry}.
Consider the homotopy
\begin{equation}
   h(x, t) = x^2 - (t - 1)^4 = (x - (t-1)^2)(x + (t-1)^2) = 0,
\end{equation}
which has the obvious two solutions $x(t) = \pm (t-1)^2$.

In this case, the power series for both paths are polynomials
of degree two, and there is no limit, as all coefficients $c_n = 0$,
for $n > 2$.  
In~\cite{PV10}, an algorithm to sweep an algebraic curve for singularities 
monitors the determinant of the Jacobian matrix along the curve.
If the path of the determinant of the Jacobian matrix on the curve
is concave up, then that is an indicator for undetected singularities.

\subsection{A Random 4-Dimensional Monomial Homotopy}

In this section, we illustrate the need for multiple precision,
even already in relatively low dimensions and degrees.
Consider
\begin{equation}
   \bfh(\x,t) =
   \left\{
\def\arraystretch{1.2}
      \begin{array}{rcl}
         x_1^7 x_2^7 x_3^7 x_4^7 & = & 1 - t \\
         x_1^7 x_2^3 x_3^2 & = & 1 - t \\
         x_2^5 x_3 x_4 & = & 1 - t \\
         x_2^7 x_3^2 x_4^2 & = &  1 - t.
      \end{array}
   \right.
\end{equation}
Storing the exponents of the monomials in the columns of
$A = [\bfa_1, \bfa_2, \bfa_3, \bfa_4]$,
$\x^A = (\x^{\bfa_1}, \x^{\bfa_2}, \x^{\bfa_3}, \x^{\bfa_4})$,
the monomial homotopy $\bfh(\x,t)$ can be written as
\begin{equation}
  \x^A = (1 - t)\bfe, \quad
  \bfe = \left[
           \begin{array}{c}
              1 \\ 1 \\ 1 \\ 1
           \end{array}
         \right], \quad
  A = \left[
        \begin{array}{cccc}
           ~7 & ~7 & ~0 & ~0~ \\
           ~7 & ~3 & ~5 & ~7~ \\
           ~7 & ~2 & ~1 & ~2~ \\
           ~7 & ~0 & ~1 & ~2~ \\
        \end{array}
     \right], \quad \det(A) = -42.
\end{equation}
At $t=0$, (1,1,1,1) is one of the 42 solutions, as $42 = |\det(A)|$,
computed via the Smith normal form of~$A$.

In double precision, extrapolating on $x_1(t)$, the extrapolation
does not get any more accurate than six decimal places.
Working with coefficients computed with 32 decimal places
running the algorithms of~\cite{BV18} (implemented in PHCpack~\cite{Ver99}),
the extrapolation gives eight decimal places of accuracy,
similarly as in the square root homotopy.

For the examples in this section, Richardson extrapolation results
in an accuracy of 8 decimal places when $n=64$ and for $n=512$,
we can locate to singularity to the full double precision.


\section{Asymptotic Expansions}

Consider the coefficient $c_n$ of $t^n$ in the Taylor series.
What happens if $n$ grows:

\begin{equation}
   \left| \frac{c_n}{c_{n+1}} \right|
   \rightarrow
   \left\{
      \begin{array}{ccl}
          |z| < 1 & : & \mbox{coefficients increase,} \\
          |z| = 1 & : & \mbox{coefficients are constant,} \\
          |z| > 1 & : & \mbox{coefficients decrease.} \\
      \end{array}
   \right.
\end{equation}
Let $x(t)$ satisfy $h(x(t), t) = 0$, then in the series for $x(t)$,
we may assume that for sufficiently large $n$, the magnitude
of the $n$th coefficient is $|z|^n$.
If we then set 
\begin{equation} \label{eqScale}
    t = |z| s
\end{equation}
then the coefficient of $s^n$ 
in the series $x(s)$ will have a magnitude close to one.
By Lemma~\ref{lemmaScale},
the radius of convergence of the series $x(s)$
equals one.

\begin{lemma} \label{lemmaScale}
Let $x(t)$ be a power series with $c_n$ as the $n$th coefficient of $t^n$
and 
\begin{equation} \label{eqClimit}
   \lim_{n \rightarrow \infty} \frac{c_n}{c_{n+1}}
   = z \in \cc \setminus \{ 0 \}.
\end{equation}
Then the series $x(t = |z| s)$ has convergence radius equal to one.
\end{lemma}
\noindent {\em Proof.}  Consider the effect of the substitution $t = |z| s$,
respectively on the $n$th and the $(n+1)$th term in the series $x(t)$:
\begin{equation}
   c_n t^n \rightarrow \underbrace{c_n |z|^n}_{=:~d_n} s^n, \quad
   c_{n+1} t^{n+1} \rightarrow 
   \underbrace{c_{n+1} |z|^{n+1}}_{=:~d_{n+1}} s^{n+1}.
\end{equation}
Then $d_n$ is the coefficient of $s^n$ in the series $x(s)$ and
\begin{displaymath}
   \left| \frac{d_n}{d_{n+1}} \right|
   = \left| \frac{c_n}{c_{n+1}} \right| \frac{1}{|z|}.
\end{displaymath}
By~(\ref{eqClimit}), 
$\displaystyle \lim_{n \rightarrow \infty}
 \left| \frac{d_n}{d_{n+1}} \right| = 1$.
Thus, $x(s)$ has a convergence radius equal to one.~\hfill~Q.E.D.

\bigskip

If interested only in the magnitude of the radius,
then in the natural application of Lemma~(\ref{eqScale}), $|z|$ is used.
Using complex arithmetic, the series $x(t = z \cdot s)$ has radius of
convergence equal to one.

\bigskip

In practice, the transformation as defined in
as defined in~(\ref{eqScale}) has numerical benefits.
In theory, it implies that without loss of generality,
we may assume that all series
we consider all have convergence radius one.

\begin{proposition} \label{propSqrt}
Assume $x(t)$ is a series which satisfies the conditions
of Theorem~\ref{thmFabry}, with a radius of convergence equal to one.
Let $c_n$ be the coefficient of $t^n$ in the series.
Then $|1 - c_n/c_{n+1}|$ is $O(1/n)$ for sufficiently large~$n$.
\end{proposition}
\noindent {\em Proof.} Expressing the Taylor series of $x(t)$ as
\begin{equation} \label{eqXseries}
   x(t) = x(0) + x'(0) t 
+ \frac{x''(0)}{2!} t^2
+ \frac{x'''(0)}{3!} t^3 + \cdots
+ \frac{x^{(n)}(0)}{n!} t^n + \cdots
\end{equation}
leads to a formula for the coefficient of $t^n$ as
\begin{equation}
   c_n = \frac{x^{(n)}(0)}{n!} \quad \mbox{and} \quad
   c_{n+1} = \frac{x^{(n+1)}(0)}{(n+1)!}.
\end{equation}
Then the error is
\begin{equation} \label{eqError}
   \left| 1 - \frac{c_n}{c_{n+1}} \right|
   = \left| 1 - \left( \frac{x^{(n)}(0)}{x^{(n+1)}(0)} \right)  (n+1) \right|
   \approx 0, \quad \mbox{for large } n.
\end{equation}

Under the assumption that the radius of convergence is equal to one,
without loss of generality we may assume that the singularity occurs
at $t = 1$.  Otherwise, if $t = z$ for some complex number~$z$,
with $|z| = 1$, we can rotate the coordinate system so $z = 1$ 
in the rotated coordinate system.
Therefore, we may assume there is a power series $p(t)$, so
\begin{equation}
   x(t) = \frac{p(t)}{1 - t} = u(t) p(t),
   \quad u(t) = \frac{1}{1 - t} = 1 + t + t^2 + t^3 + \cdots.
\end{equation}
The $p(t)/(1-t)$ can be viewed as the limit of the Pad{\'e}
approximant of degree $[n/1]$, for $n \rightarrow \infty$.
This Pad{\'e} approximant is well defined under the assumption
of Theorem~\ref{thmFabry}.
In the limit reasoning for $n \rightarrow \infty$, we work with
sufficiently large $n$, but never take $\infty$ for~$n$.

Applying Leibniz rule to the $n$th derivative of $x(t)$ leads to
\begin{equation}
   x^{(n)}(t) = \sum_{k=0}^n 
   \left( \frac{n!}{k! (n-k)!} \right) u^{(n-k)}(t) p^{(k)}(t).
\end{equation}
At $t = 0$, we have $u^{(n-k)}(0) = (n-k)!$ and we obtain
\begin{equation}
   x^{(n)}(0) = \sum_{k=0}^n \left( \frac{n!}{k!} \right) p^{(k)}(0).
\end{equation}
We rewrite the expression for $x^{(n+1)}(0)$ as
\begin{eqnarray}
   x^{(n+1)}(0) 
   & = & \sum_{k=0}^{n+1} \left( \frac{(n+1)!}{k!} \right) p^{(k)}(0) \\
   & = & \sum_{k=0}^{n} (n+1) \left( \frac{n!}{k!} \right) p^{(k)}(0)
         + p^{(n+1)}(0) \\
   & = & (n+1) x^{(n)}(0) + p^{(n+1)}(0). \label{eqXnp1zero}
\end{eqnarray}

Then we can write~(\ref{eqError}) as
\begin{eqnarray}
   \left| 1 - \frac{c_n}{c_{n+1}} \right|
   & = & \left| 1 - 
           \left( \frac{x^{(n)}(0)}{(n+1) x^{(n)}(0) + p^{(n+1)}(0)}
           \right) (n+1)
         \right| \\
   & = & \left| 1 -
           \frac{1}{1 + \frac{1}{n+1}
               \left( \frac{p^{(n+1)}(0)}{x^{(n)}(0)} \right)}
         \right|. \label{eqError2}
\end{eqnarray}
Note that we may divide by 
$x^{(n)}(0)$, because $x^{(n)}(0) \not= 0$ by the assumption
that $c_n/c_{n+1}$ is well defined for all values of~$n$,
otherwise the limit would not exist.
Denote
\begin{equation}
    C = \frac{p^{(n+1)}(0)}{x^{(n)}(0)}.
\end{equation}

Then the result follows from another series expansion:
\begin{equation} \label{eqResult}
  \frac{1}{1 + \frac{C}{n+1}}
   = 
   1 - \left( \frac{C}{n+1} \right)
     + \left( \frac{C}{n+1} \right)^2 - \cdots.
\end{equation}
Substituting the right hand side of~(\ref{eqResult})
into~(\ref{eqError2}) gives
\begin{equation} \label{eqResult2}
   \left| 1 - \frac{c_n}{c_{n+1}} \right|
   = \left|
         \left( \frac{C}{n+1} \right)
       - \left( \frac{C}{n+1} \right)^2
       + \cdots \right|
\end{equation}

What remains to prove is that $C$ does not depend on~$n$.
Dividing~(\ref{eqXnp1zero}) by $x^{(n)}(0)$ leads to
\begin{equation}
   \frac{x^{(n+1)}(0)}{x^{(n)}(0)} = n + 1 + 
   \frac{p^{(n+1)}(0)}{x^{(n)}(0)}.
\end{equation}
The assumption that $x(t)$ has a radius of convergence equal to one
implies $c_{n+1}~\approx~c_n$ and that
\begin{equation}
   \frac{x^{(n+1)}(0)}{x^{(n)}(0)} = n + O(1),
\end{equation}
and thus we have
\begin{equation}
   n + O(1) =  n + 1 + \frac{p^{(n+1)}(0)}{x^{(n)}(0)}
   \quad \mbox{or equivalently} \quad
   \frac{p^{(n+1)}(0)}{x^{(n)}(0)} \mbox{ is } O(1).
\end{equation}
Therefore $C$ is a constant, independently of~$n$.
This shows that the error is $O(1/(n+1))$.
For large~$n$, $O(1/(n+1))$ is $O(1/n)$. \hfill Q.E.D.

\bigskip

Observe that the above proof does not make any assumptions
on the type of homotopy used, other than the existence of a limit
as in the theorem of Fabry.
Then the main result of this section can be stated as below.

\begin{corollary} \label{corRichardson}
Assuming the convergence radius equals one,
applying Richardson extrapolation $N$ times on a Taylor series
truncated after $n$ terms, results in an $O(1/n^{N+1})$ error
on the radius of convergence.
\end{corollary}
{\em Proof.} By Proposition~\ref{propSqrt}, 
and in particular the expansion in~(\ref{eqResult2}), we have
\begin{equation} \label{eqErrorExpansion1}
   1 + \gamma_1 \left( \frac{1}{n} \right) 
     + \gamma_2 \left( \frac{1}{n} \right)^2 
     + \gamma_3 \left( \frac{1}{n} \right)^3 + \cdots
\end{equation}
as the expansion for the error to the limit~1.

For $N=1$, the first extrapolated values have error~$O(1/n^2)$,
because the leading terms of the errors are $O(1/n)$
and running Richardson extrapolation once 
(for $j=2$ and $i=2,3,\ldots,N$ in~(\ref{eqRichardson}))
eliminates this leading term.

Using the formulas in~(\ref{eqRichardson}) to compute 
the next columns in the triangular table eliminates the next terms 
in the error expansion in~(\ref{eqErrorExpansion1}).
After extrapolating $N-1$ more times, 
we then obtain an~$O(1/n^{N+1})$ error term. \hfill Q.E.D.

\bigskip

The assumption that the radius of convergence equals one
makes the Richardson extrapolation superfluous,
as the outcome of the extrapolation is already known.
We can remove this assumption.  
Consider for example the homotopy 
$h(x,t) = x^2 - 2 + t = 0$
and $x(t) = \sqrt{2 - t}$ as the positive solution branch.
If $c_n$ is the $n$th coefficient of the Taylor series, then
\begin{equation}
   \frac{c_n}{c_{n+1}} = 2 \left( \frac{2 (n+1)}{2n - 1} \right)
                       = 2 f(n),
\end{equation}
where $f(n)$ is the formula from~(\ref{eqSqrtFabry}).
Similarly, for the homotopy $h(x,t) = x^2 - 1/2 + t = 0$
and $x(t) = \sqrt{1/2 - t}$ as the positive solution branch,
with $c_n$ as the $n$th coefficient of the Taylor series, we have
\begin{equation}
   \frac{c_n}{c_{n+1}} = \frac{1}{2} \left( \frac{2 (n+1)}{2n - 1} \right)
                       = \frac{1}{2} f(n).
\end{equation}
This implies that for those two examples, the series development
of $f(n)$ in $1/n$ is multiplied respectively with 2 or 1/2,
and that therefore Richardson extrapolation applies.

\begin{theorem} \label{thmRichardson}
Let $c_n$ be the coefficient with $t^n$ in $x(t)$
and denote $f(n) = c_n/c_{n+1}$.  If
\begin{equation}
   \lim_{n \rightarrow \infty} \frac{c_n}{c_{n+1}}
   = z \in \cc \setminus \{ 0 \},
\end{equation}
then
\begin{equation} \label{eqZseries}
   f(n) = z
   + \gamma_1 z \left( \frac{1}{n} \right) 
   + \gamma_2 z \left( \frac{1}{n} \right)^2 
   + \gamma_3 z \left( \frac{1}{n} \right)^3 + \cdots. 
\end{equation}
\end{theorem}
{\em Proof.}  By Lemma~\ref{lemmaScale}, we transform $x(t)$
into $x(s) = x(t = z \cdot s)$, which has convergence radius one.
Let $d_n$ be the coefficient of $s^n$ in $x(s)$
and denote $g(n)~=~d_n/d_{n+1}$.
For $g(n)$, we have the expansion~(\ref{eqErrorExpansion1}):
\begin{equation} \label{eqErrorExpansion2}
   g(n) = 1
   + \gamma_1 \left( \frac{1}{n} \right) 
   + \gamma_2 \left( \frac{1}{n} \right)^2 
   + \gamma_3 \left( \frac{1}{n} \right)^3 + \cdots.
\end{equation}
The above series development is unique.
Therefore, transforming $s = t/z$, 
gives the series~(\ref{eqZseries}).~\hfill~Q.E.D.

\bigskip

Theorem~\ref{thmRichardson} provides the justification
for the application of Richardson extrapolation
and the statement of Corollary~\ref{corRichardson}
holds in theory for any series, not only for those
with radius of convergence equal to one.
However, in practice, series with a radius of convergence
smaller than one will have very large coefficients
which cause numerical instabilities 
and unavoidably arithmetical overflow.

If the convergence radius of a power series equals one,
then it is safe to calculate the coefficients of the power series
from sample points at nearby locations.

\section{Fourier Series}

In computational complex analysis~\cite{Hen79},
the discrete Fourier transform is applied to compute
the coefficients of the Taylor series.
For general references on the application of Fourier transforms
in computer algebra and numerical analysis, 
we refer to~\cite{vGG99} and~\cite{CF13}.

\begin{figure}[hbt]
\begin{center}
\begin{tikzpicture}
\draw (0cm, 0cm) circle(8mm);
\draw (0cm, 0cm) circle(15mm);
\draw[-{Latex[length=2mm, width=1mm]}] (0mm, 0mm) -- (15mm, 0mm);
\draw[-{Latex[length=2mm, width=1mm]}] (0mm, 0mm) -- (0mm, 8mm);
\node[text width=5mm] at (0mm, 3mm) {\scriptsize $h$};
\node[text width=5mm] at (12mm, -2mm) {\scriptsize $r$};
\end{tikzpicture}
\caption{The radius of convergence~$r$ and step size~$h$.  We want $h \ll r$.}
\label{figstepradius}
\end{center}
\end{figure}
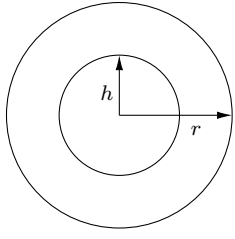

As described in~\cite{Nas16}, many derivatives are computed
simultaneously with an accuracy close to machine precision,
for a suitable step size, using complex arithmetic,
extending the complex-step differentiation method~\cite{ST98}
to higher order derivatives.
Figure~\ref{figstepradius} illustrates the problem:
the step size must be smaller than the radius of convergence.
This problem is addressed in section~\ref{secReconditioning}.

To introduce the application of the discrete Fourier transform to
compute the Taylor series, consider the development of $f$ at $z$,
using step size $h \omega$:
\begin{eqnarray}
   f(z + h \omega)
 & = & f(z) + h \omega f'(z) + \frac{h^2}{2} \omega^2 f''(z)
   + \frac{h^3}{3!} \omega^3 f'''(z) \\
 & + & \frac{h^4}{4!} \omega^4 f^{(\textup{\romannumeral 4})}(z)
   +  \frac{h^5}{5!} \omega^5 f^{(\textup{\romannumeral 5})}(z)
   + \frac{h^6}{6!} \omega^6 f^{(\textup{\romannumeral 6})}(z) \\
 & + & \frac{h^7}{7!} \omega^7 f^{(\textup{\romannumeral 7})}(z)
   + \frac{h^8}{8!} \omega^8 f^{(\textup{\romannumeral 8})}(z) + \cdots,
\end{eqnarray}
where $\omega$ is the eight complex root of unity: $\omega^8 = 1$.
Regrouping in powers of $\omega$ then gives
\begin{eqnarray}
   f(z + h \omega)
  & = & f(z) 
+ \frac{h^8}{8!} f^{(\textup{\romannumeral 8})}(z) + \cdots \\
  & + & \omega \left( h f'(z)
+ \frac{h^9}{9!} f^{(\textup{\romannumeral 9})}(z) + \cdots \right) \\
  & + & \omega^2 \left( \frac{h^2}{2!} f''(z)
+ \frac{h^{10}}{10!} f^{(\textup{\romannumeral 10})}(z) + \cdots \right) \\
  & + & \omega^3 \left( \frac{h^3}{3!} f'''(z)
+ \frac{h^{11}}{11!} f^{(\textup{\romannumeral 11})}(z) + \cdots \right) \\
  & + & \omega^4 \left( \frac{h^4}{4!} f^{(\textup{\romannumeral 4})}(z) 
  + \frac{h^{12}}{12!} f^{(\textup{\romannumeral 12})}(z) + \cdots \right) \\
  & + & \omega^5 \left( \frac{h^5}{5!} f^{(\textup{\romannumeral 5})}(z) 
  + \frac{h^{13}}{13!} f^{(\textup{\romannumeral 13})}(z) + \cdots \right) \\
  & + & \omega^6 \left( \frac{h^6}{6!} f^{(\textup{\romannumeral 6})}(z) 
  + \frac{h^{14}}{14!} f^{(\textup{\romannumeral 14})}(z) + \cdots \right) \\
  & + & \omega^7 \left( \frac{h^7}{7!} f^{(\textup{\romannumeral 7})}(z) 
  + \frac{h^{15}}{15!} f^{(\textup{\romannumeral 15})}(z) + \cdots \right).
\end{eqnarray}
For $k$ from 1 to 7, the coefficients of $\omega^k$ allow the extraction
of the $k$th derivative of $f$ at $z$, at a precision of $O(h^8)$.

The Discrete Fourier Transform
\begin{equation}
   \begin{array}{ccccc}
      \mbox{DFT}_\omega & : & \cc^n & \rightarrow & \cc^n \\
      & & (f_0, f_1, \ldots, f_{n-1}) & \mapsto &
      (F(\omega^0), F(\omega^1), \ldots, F(\omega^{n-1}))
   \end{array}
\end{equation}
takes the coefficients of the polynomial $F$ with 
coefficients $f_0$, $f_1$, $\ldots$, $f_{n-1}$,
where $\omega^n = 1$ and returns the values of $F$
at the powers of~$\omega$.
The inverse of $\mbox{DFT}_\omega$ returns the coefficients
of $\omega^k$ needed in the Taylor series of $f(z + h \omega)$.

As illustrated by Table~\ref{tabSqrtDiff}, the derivatives grow
as fast as $n!$ and therefore, except for small~$n$, 
we may not expect to obtain highly accurate values.

\begin{table}[hbt]
\begin{center}
\caption{Derivatives of $x(t) = \sqrt{1 - t}$ at $t = 0$.
The approximate values are computed with step size $h = 0.5$.
The last column is the relative error.}
\begin{tabular}{r||r|r|r}
$n$ & \multicolumn{1}{c|}{exact $x^{(n)}(0)$}
    & \multicolumn{1}{c|}{approximation $x^{(n)}(0)$}
    & \multicolumn{1}{c}{error} \\ \hline \hline
 0  &             $1.000000000000$  &             $0.999999968596$  &  $3.14\mbox{E}\!-\!08$ \\
 1  &            $-0.500000000000$  &            $-0.500000028787$  &  $5.76\mbox{E}\!-\!08$ \\
 2  &            $-0.250000000000$  &            $-0.250000053029$  &  $2.12\mbox{E}\!-\!07$ \\
 3  &            $-0.375000000000$  &            $-0.375000147155$  &  $3.92\mbox{E}\!-\!07$ \\
 4  &            $-0.937500000000$  &            $-0.937500546575$  &  $5.83\mbox{E}\!-\!07$ \\
 5  &            $-3.281250000000$  &            $-3.281252546540$  &  $7.76\mbox{E}\!-\!07$ \\
 6  &           $-14.765625000000$  &           $-14.765639282757$  &  $9.67\mbox{E}\!-\!07$ \\
 7  &           $-81.210937500000$  &           $-81.211031230822$  &  $1.15\mbox{E}\!-\!06$ \\
 8  &          $-527.871093750000$  &          $-527.871798600561$  &  $1.34\mbox{E}\!-\!06$ \\
 9  &         $-3959.033203125000$  &         $-3959.039180858922$  &  $1.51\mbox{E}\!-\!06$ \\
10  &        $-33651.782226562500$  &        $-33651.838679975779$  &  $1.68\mbox{E}\!-\!06$ \\
11  &       $-319691.931152343750$  &       $-319692.518875707698$  &  $1.84\mbox{E}\!-\!06$ \\
12  &      $-3356765.277099609375$  &      $-3356771.966430745088$  &  $1.99\mbox{E}\!-\!06$ \\
13  &     $-38602800.686645507812$  &     $-38602883.297614447773$  &  $2.14\mbox{E}\!-\!06$ \\
14  &    $-482535008.583068847656$  &    $-482536106.155545711517$  &  $2.27\mbox{E}\!-\!06$ \\
15  &   $-6514222615.871429443359$  &   $-6514238371.741491317749$  &  $2.42\mbox{E}\!-\!06$ \\
16  &  $-94456227930.135726928711$  &  $-94456466497.677398681641$  &  $2.53\mbox{E}\!-\!06$ \\
\end{tabular}
\label{tabSqrtDiff}
\end{center}
\end{table}
The step size of $h=0.5$ used in Table~\ref{tabSqrtDiff} is a compromise value.
Values of $h$ smaller than $0.5$ give more accurate results
for the lower order derivatives but give then
too inaccurate values for the higher order derivatives.
The opposite happens for values of $h$ larger than~0.5.

Fortunately, we do not need the derivatives $x^{(n)}(0)$, but the
coefficients of the Taylor series, $c_n = x^{(n)}(0)/n!$.
Table~\ref{tabSqrtSeries} shows the application of the DFT to
compute the series coefficients.  Compared to the derivatives
in Table~\ref{tabSqrtDiff}, the computations in double precision
arithmetic give six decimal places of accuracy for $n=64$.
The step size $h = 0.85$ gave the most accurate results.

\begin{table}[hbt]
\begin{center}
\caption{Coefficients $c_n$ of the Taylor series of $x(t) = \sqrt{1 - t}$ at $t = 0$.
The approximate values are computed with step size $h = 0.85$.
The last column is the relative error.}
\begin{tabular}{r||r|r|r}
$n$ & \multicolumn{1}{c|}{exact $c_n$}
    & \multicolumn{1}{c|}{approximation}
    & \multicolumn{1}{c}{error} \\ \hline \hline
 0  &  $ 1.000000000000$  &  $ 0.999999986011$  &  $1.40\mbox{E}\!-\!08$ \\
 1  &  $-0.500000000000$  &  $-0.500000013671$  &  $2.73\mbox{E}\!-\!08$ \\
 2  &  $-0.125000000000$  &  $-0.125000013365$  &  $1.07\mbox{E}\!-\!07$ \\
 4  &  $-0.039062500000$  &  $-0.039062512786$  &  $3.27\mbox{E}\!-\!07$ \\
 8  &  $-0.013092041016$  &  $-0.013092052762$  &  $8.97\mbox{E}\!-\!07$ \\
32  &  $-0.001576932599$  &  $-0.001576940258$  &  $4.86\mbox{E}\!-\!06$ \\
64  &  $-0.000554221198$  &  $-0.000554226120$  &  $8.88\mbox{E}\!-\!06$ \\
\end{tabular}
\label{tabSqrtSeries}
\end{center}
\end{table}

In machine double precision,
the results in Table~\ref{tabSqrtDiff} and Table~\ref{tabSqrtSeries}
are close to optimal, with the step sizes respectively equal
to~$0.5$ and~$0.85$.  Using those large step sizes in multiprecision
will not give more accurate results, but multiprecision will allow
to select smaller step sizes.  In particular with 33 decimal places
(using mpmath 1.1.0~\cite{mpmath} with SymPy~1.4~\cite{SymPy11}
in Python~3.7.3), the 16th derivative is computed with an accuracy
of 15 decimal places, with step size~0.1 and 
the error on the 64th coefficient coefficient on the series
drops to $10^{-11}$, with step size~0.5.

Instead of working with the same step size for all series coefficients,
alternatively, one could explore using different step sizes.
In this context, one classical and very common application
of Richardson extrapolation is 
to improve the accuracy of numerical differentiation.

When $z$ is a complex number,
the complex step derivative is generalized in~\cite{Kim21} 
and~\cite{RDH21} with quaternion arithmetic.
Using the quaternion Fourier transform~\cite{ELS14},~\cite{SLS08},
the coefficients of the Taylor series can be computed.

\section{Polynomial Homotopies}

The homotopies in this section have multiple singularities
in the complex plane, for complex values of $t$,
with real part $< 1$, but only one singularity at $t = 1$.
Knowing the location of the last pole leads to the reconditioning
of the homotopy and to series with convergence radius equal to one.

\subsection{The Last Pole}

Let $\bff(\x) = \zero$ be the system we want to solve
and assume we have at least one solution of~$\bfg(\x) = \zero$.
Then the homotopy

\begin{equation} \label{eqGammaHomotopy}
   \bfh(\x,t) = \gamma (1-t) \bfg(\x) + t \bff(\x) = \zero,
   \quad t \in [0,1], \quad \gamma \in \cc, |\gamma| = 1,
\end{equation}
defines a path starting at $t = 0$, at a solution of $\bfg(\x) = \zero$
and ending at $t = 1$, at a solution of $\bff(\x) = \zero$.
The constant $\gamma$ is a random complex number.
If $\bfg(\x) = \zero$ has no singular solutions,
then it follows from the main theorem of elimination theory
that all paths defined by $\bfh(\x, t) = \zero$ are regular and bounded
for $t \in [0, 1)$, except for finitely many {\em complex} values for~$t$.
In~\cite{SVW05}, this constructive argument is illustrated 
by examples of homotopies of small degrees and dimension.

The key point is the existence of a polynomial $H(t)$ of finite degree,
with $H(0) \not= 0$, as $\bfg(\x) = \zero$ has no singular solutions.
Moreover, by the random {\em complex} choice of $\gamma$,
all roots of $H$ are in the complex plane, except for $t = 1$,
if the system $\bff(\x) = \zero$ has a singular solution.
By construction of $\bfh(\x,t) = \zero$, we can introduce 
the notion of {\em the last pole}, as the complex number~$\rho$,
for which $H(\rho) = 0$ and of all roots of $H$, $\rho$ has
the largest real part less than one.\footnote{If all real parts of the
roots of $H$ are larger than one, then we are in the case similar
to a monomial homotopy, a case that is then already solved.}

\begin{figure}[hbt]
\begin{center}
\begin{picture}(330,60)(0,0)
\put(0,0){
\begin{tikzpicture}
\draw (0cm, 0cm) circle(1cm);
\draw (0mm, -2mm) -- (0mm, 2mm);
\draw (-7mm, -2mm) -- (-7mm, 2mm);
\draw (-7mm, 0mm) -- (-5mm, 8.66mm);
\draw[fill=black] (1cm, 0cm) circle(0.7mm);
\draw (-5mm, 8.66mm) circle(0.7mm);
\draw[-{Latex[length=2mm, width=1mm]}] (-15mm, 0mm) -- (17mm, 0mm);
\node[text width=5mm] at (-6mm, 10mm) {$\rho$};
\node[text width=5mm] at (20mm, -1mm) {$t$};
\node[text width=5mm] at (13mm, 2mm) {$1$};
\node[text width=10mm] at (-3mm, -4mm) {$t_0 < t_*$};
\end{tikzpicture}
}
\put(110,0){
\begin{tikzpicture}
\draw (0cm, 0cm) circle(1cm);
\draw (0mm, -2mm) -- (0mm, 2mm);
\draw (0mm, 0mm) -- (-5mm, 8.66mm);
\draw[fill=black] (1cm, 0cm) circle(0.7mm);
\draw (-5mm, 8.66mm) circle(0.7mm);
\draw[-{Latex[length=2mm, width=1mm]}] (-15mm, 0mm) -- (17mm, 0mm);
\node[text width=5mm] at (-6mm, 10mm) {$\rho$};
\node[text width=5mm] at (20mm, -1mm) {$t$};
\node[text width=5mm] at (13mm, 2mm) {$1$};
\node[text width=10mm] at (0mm, -4mm) {$t_0 = t_*$};
\end{tikzpicture}
}
\put(220,0){
\begin{tikzpicture}
\draw (0cm, 0cm) circle(1cm);
\draw (0mm, -2mm) -- (0mm, 2mm);
\draw (7mm, -2mm) -- (7mm, 2mm);
\draw (7mm, 0mm) -- (-5mm, 8.66mm);
\draw[fill=black] (1cm, 0cm) circle(0.7mm);
\draw (-5mm, 8.66mm) circle(0.7mm);
\draw[-{Latex[length=2mm, width=1mm]}] (-15mm, 0mm) -- (17mm, 0mm);
\node[text width=5mm] at (-6mm, 10mm) {$\rho$};
\node[text width=5mm] at (20mm, -1mm) {$t$};
\node[text width=5mm] at (13mm, 2mm) {$1$};
\node[text width=10mm] at (4mm, -4mm) {$t_* < t_0$};
\end{tikzpicture}
}
\end{picture}
\caption{Schematic of the last pole $\rho$ marked by the hollow circle.
At the center, at $t = t_0 = t_*$, $\rho$ and $1$ are at the
same distance.  At $t_0 < t_*$, the proximity of $\rho$ determines
the step size, while for $t_* < t_0$, the singularity at one will
be detected.}
\label{figlastpole}
\end{center}
\end{figure}
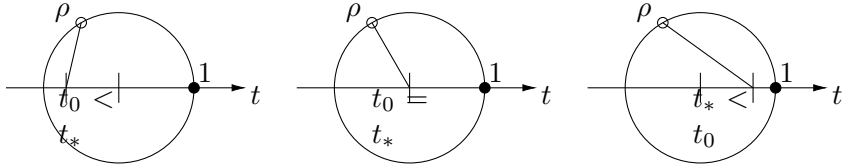

Figure~\ref{figlastpole} illustrates that $\rho$ is the last
complex singular value detected by the radar of a path tracker
which applies the theorem of Fabry to set its step size.

By construction of the homotopy~$\bfh(\x, t) = \zero$,
and in particular by the random choice of the complex constant $\gamma$,
the solution at $t = t_*$ is regular, and well conditioned.
This implies that Newton's method for the series coefficients
converges quadratically.  
One could then already discover the singular solution for $t = 1$,
via the computation of a Pad{\'{e}} approximant with quadratic denominator.
Via a perturbation argument, for $t = t_* + \delta$,
for suitable $\delta > 0$, the application
of the theorem of Fabry will detect $t = 1$ as a singular solution,
{\em without computing~$\x(t)$ for $t \approx 1$.}

\subsection{Homotopy Reconditioning} \label{secReconditioning}

Once the path tracker reaches a value for the continuation parameter $t$,
that is past the last pole towards an isolated singularity
at $t=1$, at the end of a path, the coefficients of the power series
will grow very fast, which is already an indication for the trouble to come.

For the reliable numerical computation of the power series,
consider the transformation in the homotopy $\bfh(\x, t) = \zero$:
\begin{equation} \label{eqReconditioning}
   t = r s + t_0, \quad r = 1 - t_0, \quad t_0 = t_* + \delta,
\end{equation}
where $t_*$ is value as in Figure~\ref{figlastpole},
at the same distance from the last pole $\rho$ and the end point $t = 1$,
and $\delta$ is a suitable positive value so at $t_0 = t_* + \delta$,
the application of the theorem of Fabry will detect $t = 1$ as the
location for the closest singular solution.

After applying~(\ref{eqReconditioning}), the series development
of the path $\x(s)$ defined by the homotopy $\bfh(\x(s), s) = \zero$
will have convergence radius equal to one.
The term {\em reconditioning} is justified as the coefficients of
the Taylor series in the reconditioned homotopy 
do not grow exponentially fast.

\section{Computational Experiments}

The new methods are illustrated with computational experiments
on two well known examples in the literature, with ad hoc tools,
using test procedures in version 2.4.85 PHCpack~\cite{Ver99}
(with QDlib~\cite{HLB01} and CAMPARY~\cite{JMPT16} for
multiple double arithmetic),
version 1.1.1 of phcpy~\cite{OFV19}, 
version 1.4 of sympy~\cite{SymPy11}, and 
version 1.1.0 of mpmath~\cite{mpmath}, in Python~3.7.3.
The computations were done on a CentOS~6.10 Linux computer
with 23.4GB of memory and a 12-core Intel Xeon X5690 at~3.47Ghz.

\subsection{Ojika's First Example}

One example in~\cite{Oji87a}
(known in benchmarks as {\tt ojika1},
used in~\cite{HJLZ20}, \cite{LVZ06}, \cite{LZ12,LZ14}) is
\begin{equation}
   \bff(x,y) =
   \left\{
     \begin{array}{rcl}
        x^2 + y - 3 & = & 0 \\
        x + 0.125 y^2 - 1.5 & = & 0.
     \end{array}
   \right.
\end{equation}
This system has one regular solution at $(-3, 6)$
and a triple root at~$(1, 2)$.
Using \newline $\gamma = -0.917153159675641 - 0.398534919043474~\!I$,
$I = \sqrt{-1}$,
in the homotopy~(\ref{eqGammaHomotopy}) with start system
\begin{equation}
   \bfg(x,y) =
   \left\{
     \begin{array}{rcl}
        x^2 - 1 & = & 0 \\
        y^2 - 1 & = & 0 \\
     \end{array}
   \right.
\end{equation}
makes that the path starting at $(1,1)$ converges
to the triple root.

The value $t_0$ after $t_*$ (the location of the last pole) 
that was used is $t_0 = 0.955647336181678$.
At this value for $t$, the coordinates of the corresponding solution are
\begin{eqnarray} 
  x & \approx & 1.17998166418735 + 0.0181391513338172~\!I, \\
  y & \approx & 1.60871001974391 - 0.0423866308603763~\!I,
\end{eqnarray}
with the inverse of the condition number estimated at {\tt 8.9E-03}.
In double precision, a condition number of about $10^3$ 
is within the range of what is considered well conditioned.
Observe that the coordinates of the solution corresponding to~$t_0$
are far from the location of the triple root~$(1, 2)$.

The value for $r = 1 - t_0$ is $0.044352663818322036$,
which implies that, without reconditioning, the magnitude of
the Taylor series coefficients will increase with about
two decimal places.  At that pace, as $2^{64} \approx 10^{+19}$,
numerical difficulties arise without reconditioning.

After reconditioning, with $n = 64$, the ratio, 
based on the power series for the first coordinate $x(s)$,
is estimated at
\begin{equation}
   1.0265192231142901 + 2.9197227799819557\mbox{E}\!-\!05~\!I
\end{equation}
and the magnitude of the imaginary part corresponds to the
magnitude of the coefficients $c_n$ in the series of~$x(s)$.
This mild decline of the exponents corresponds to the
over estimation of the radius at about~$1.0265$.
Applying Richardson extrapolation yields
\begin{equation}
   0.9999729580138075 + 8.484367218447337\mbox{E}\!-\!06~\!I,
\end{equation}
which thus locates the singularity with an error of $10^{-6}$.

The above computations were done in double precision.
In double double precision ($\approx$ 32 decimal places), with $n = 512$,
the ratio is first estimated at $1.00326$
and Richardson extrapolation then improves the accuracy,
to obtain an error of $10^{-6}$ on the value $t = 1$, 
the location of the singularity,
confirming the result obtained in double precision.

\subsection{One Fourfold Root of Cyclic 9-roots}

The cyclic $n$-roots problem
\begin{equation} \label{eqcyclicsys}
  \bff(\x) =
  \left\{
   \begin{array}{c}
   x_{0}+x_{1}+ \cdots +x_{n-1}=0 \\
   i = 2, 3, \ldots, n-1: 
    \displaystyle\sum_{j=0}^{n-1} ~ \prod_{k=j}^{j+i-1}
    x_{k~{\rm mod}~n}=0 \\
   x_{0}x_{1}x_{2} \cdots x_{n-1} - 1 = 0 \\
  \end{array}
  \right.
\end{equation}
is a well known benchmark problem in polynomial system solving,
which arose in the study of biunimodular vectors~\cite{FR15}.
The cyclic 9-roots problem was solved in~\cite{Fau01},
and its roots of multiplicity four were used in the development
of deflation in~\cite{LVZ06}.
This system was used to illustrate the computation
of the multiplicity structure in~\cite{DZ05}.

The start system $\bfg(\x) = \zero$ in a homotopy
to solve $\bff(\x) = \zero$ was obtained by running the plain blackbox 
solver (the extended version is described in~\cite{Ver18})
on 12 cores tracking 11,016 is less than two minutes.  
For reproducibility, the seed in the random number generators
was~$7131$.  That $\bfg(\x)$ was then used in 
the homotopy~(\ref{eqGammaHomotopy}) with
$\gamma = -0.917153159675641 - 0.398534919043474~\!I$, $I = \sqrt{-1}$.
One path was selected that ended at one of the fourfold roots.

The value for $t$ after $t_*$, the location of the last pole 
is $t_0 = 0.998315512784621$, 
with coordinates of the corresponding solution 
\begin{eqnarray}
 x_0 & \approx & +1.00000126517819\hphantom{2} + 2.90396442439194\mbox{E}\!-\!07 \\
 x_1 & \approx & -2.61867609654276\hphantom{2} - 2.06312686218454\mbox{E}\!-\!03 \\
 x_2 & \approx & -0.381725080860952 + 6.25420054941098\mbox{E}\!-\!05 \\
 x_3 & \approx & +1.00151501674915\hphantom{2} + 1.11189386260303\mbox{E}\!-\!03 \\
 x_4 & \approx & +0.381629266681896 - 3.62839287359460\mbox{E}\!-\!04 \\
 x_5 & \approx & +2.62034316800711\hphantom{2} + 2.49236777820171\mbox{E}\!-\!03 \\
 x_6 & \approx & +0.998483898493147 - 1.11096857563447\mbox{E}\!-\!03 \\
 x_7 & \approx & -2.61970187995193\hphantom{2} - 4.30339092688366\mbox{E}\!-\!04 \\
 x_8 & \approx & -0.381870388949536 + 3.01610075581641\mbox{E}\!-\!04
\end{eqnarray}
with inverse condition number estimated at {\tt 5.3E-5}.
Although the homotopy does not respect the permutation symmetry,
the orbit structure of the solution can already be observed,
at the limited accuracy of about three decimal places.

The value for $r = 1 - t_0$ is $0.0016844872153789492$
and without reconditioning the homotopy,
the coefficients in the power series expansions of the solution
increase at a very high pace.
After reconditioning, with $n= 32$, the convergence radius is
estimated at
\begin{equation}
   1.00000000099639 + 4.319265\mbox{E}\!-\!09~\!I
\end{equation}
and confirmed in double double precision.
Because of the close proximity to the singularity,
no extrapolation is necessary in this case.


\vspace{-2mm}

\section{Conclusions}

Richardson extrapolation is effective to locate the closest singularity
as shown by the asymptotic expansions on the ratio of two consecutive
coefficients in the Taylor series of the solution curves,
under the condition of the theorem of Fabry.

\vspace{-1mm}

The homotopy continuation parameter can always be adjusted so
the convergence radius of the power series equals one,
which allows for a safe step size selection in the application
of the discrete Fourier transform to compute all coefficients
of the series efficiently and accurately.

\vspace{-1mm}

Deflation restores the quadratic convergence of Newton's method
on an isolated singular solution via reconditioning.
The homotopy reconditiong using the location of the last pole provides
an apriori justification for the application of the deflation method
via the Richardson extrapolation on the ratios of the coefficients
of power series.

\vspace{-1mm}

The theorem of Fabry provides a radar to detect singularities.
In this paper we have shown that this radar can accurately
locate the nearest singular solution of a polynomial homotopy.
We apply this radar at a safe distance from singularities, 
at a regular solution where the quadratic convergence
of Newton's method holds.

\noindent {\bf Acknowledgements.} 
Some of the results in this paper were presented by the first author
on 27 March 2022 in a preliminary report at the special session 
on Optimization, Complexity, and Real Algebraic Geometry,
which took place online.
The authors thank the organizers, Saugata Basu and Ali Mohammad Nezhad,
for their invitation.
We thank the three reviewers of this paper for their useful comments
which helped to improve the exposition.


\bibliographystyle{plain}

\end{document}